\documentclass{ws-p10x7-mod1}

\begin{document}

\title{Combined recoil and threshold resummation for hard scattering cross sections}

\author{Eric Laenen}

\address{NIKHEF Theory Group, Kruislaan 409\\ 1098 SJ Amsterdam, The
Netherlands}

\author{George Sterman}

\address{C.N.\ Yang Institute for Theoretical Physics,
SUNY Stony Brook\\
Stony Brook, New York 11794 -- 3840, U.S.A.\\
\ \\
Physics Department, Brookhaven National Laboratory,\\
Upton, NY 11973, U.S.A.}

\author{Werner Vogelsang}

\address{RIKEN-BNL Research Center, Brookhaven National Laboratory,\\
Upton, NY 11973, U.S.A.}

\twocolumn[\maketitle\abstract{
We discuss the simultaneous resummation of 
threshold and recoil enhancements to partonic
cross sections due to soft radiation.
Our method is based on a refactorization
of the parton cross section near its partonic
threshold. It avoids double counting, 
conserves the flow of partonic energy, and reproduces
either threshold or recoil resummation when the other 
enhancements are neglected.}]

\section{Introduction}
\label{sec:Introduction}

A large class of hard-scattering cross
sections in QCD may be
factorized into convolutions of parton distributions and
fragmentation functions with hard-scattering
functions.~\cite{cssrv} The latter are computed in
perturbation theory. Intermediate infinities associated with 
virtual and emitted soft gluons cancel in their
higher-order corrections, but finite remnants,
assuming the form of plus-distributions and delta-functions,
may lead to large enhancements when integrated against
the smooth functions in the  convolutions.
In physical terms, soft gluon radiation affects 
the hard-scattering process 
by reweighting the relative importance 
of near-threshold partonic subprocesses
to the physical cross section, and by providing
the final state with overall transverse momentum $Q_T$ through its recoil.
When the conservation of energy is taken into account,
the sign of the cumulative effect is {\it a priori} unclear:
enhancement effects of recoil may be
offset by the suppression due to the extra energy 
required to produce the desired final state plus recoil radiation.
A combined and consistent analysis is required \cite{Liunified} 
and in what follows we sketch our approach to this 
problem~\cite{lsv,lsvlong}.\footnote{Contributed to the 30th International Conference on 
High-Energy Physics (ICHEP 2000), Osaka, Japan, 27 Jul - 2 Aug 2000}

\section{Electroweak annihilation}
\label{sec:Electr-annih}

These processes are
characterized at lowest order
by $ab \rightarrow V$, with $a,b$ partons
and $V$ an electroweak final state of mass $Q$
and transverse momentum ${\bf Q}_T$.
Because of their relative simplicity and phenomenological
interest, these processes
have been studied intensely for resummation purposes.
Partonic threshold is at $z\equiv Q^2/\hat{s}=1$ and ${\bf Q}_T=0$,
and the singular functions to be resummed
are plus-distributions in $1-z$ and $Q_T/Q$. 
Our method generalizes that of Ref.~\cite{dyresum}, and organizes these
distributions in the cross section.
This is done
by refactorizing partonic cross sections into short-distance
functions, $\sigma^{(H)}_{ab}$, sensitive only to the 
hardest scale, $Q$, and matrix elements of appropriate operators
that absorb all dependence on $(1-z)Q$ and $Q_T$, to 
leading power in $1-z$.
\begin{eqnarray}
  \label{eq:1}
  { d \sigma_{ab\to V} \over dQ^2 d^2{\bf Q}_T}
&=&
{1\over S}\; \sigma_{ab\to V}^{(H)}(Q^2,\alpha_s(Q^2))
\nonumber \\
&\ & \hspace{-15mm} \times
\int dx_a d^2{\bf k}_a\, R_{a/a}(x_a,{\bf k}_a,Q)
\nonumber \\
&\ & \hspace{-15mm} \times
\int dx_b d^2{\bf k}_b\, R_{b/b}(x_b,{\bf k}_b,Q)
\nonumber \\
&\ & \hspace{-15mm} \times
\int dw_s d^2{\bf k}_s\, U_{ab}(w_s,Q,{\bf k}_s)
\nonumber \\
&\ & \hspace{-15mm} \times
\delta(1-Q^2/S-(1-x_a)-(1-x_b)-w_s)
\nonumber \\
&\ & \hspace{-15mm} \times
\; \delta^2 \left({\bf Q}_T- {\bf k}_a-{\bf k}_b-{\bf k}_s\right)
+Y_{\rm j}\, ,
\end{eqnarray}
where
\begin{eqnarray}
  \label{eq:2}
  R_{f/f}(x,{\bf k},2p_0)
&=&  \nonumber\\
&\ & \hspace{-30mm}
\frac{1}{4\sqrt{2}N_c} \int
{d\lambda\over 2\pi}\; {d^2{\bf b} \over (2\pi)^2}  
{\rm e}^{-i\lambda xp_0+i{\bf b}\cdot {\bf
k}}\;   \nonumber
 \\
&\ & \hspace{-25mm}
\times\langle f(p)| \bar q_f\big(\lambda^+,{\bf b},0^-\big) \gamma^+
q_f(0) |f(p)\rangle\,
\end{eqnarray}
is a partonic quark density at fixed energy 
and transverse momentum,
and $U_{ab}$ a purely eikonal function, depending
on parton velocities $\beta$, and on the soft
radiation's energy $w_s\,Q$. The $x$'s and
${\bf k}$'s are defined by (\ref{eq:2}).
The delta functions in the triple convolution
of (\ref{eq:1}) relate the singular behavior 
of the various functions, and $Y_j$ represents
the matching to finite order.
The all-order behavior of the above functions
and the cross section
can be analyzed through their eikonalized equivalents.
The eikonal cross section can be expressed as 
an exponent, $E_{ab}$, an integral over 
diagrammatically-defined functions referred to as 
``webs'',~\cite{gath} $w_{ab}$,
\begin{eqnarray}
  \label{eq:5}
E_{ab}
&=& 2\, \int^Q {d^{4-2\epsilon}k\over
\Omega_{1-2\epsilon}}\;
\nonumber\\
&\ & \times\
w_{ab} \left(k^2,{k\cdot \beta k\cdot \beta'\over
\beta\cdot\beta'},\mu^2,\alpha_s(\mu^2),\epsilon\right)
\nonumber\\
&\ & \times
\left( {\rm e}^{-N(k_0/Q)-i{\bf k}\cdot {\bf b}}\;-1\right)\, .
\end{eqnarray}
The integral is over the 
energy and transverse momentum
contributed by each web to the
final state. The $k$-dependence of $w_{ab}$ 
follows from the invariance of the eikonal cross section
under rescalings of $\beta$ and $\beta'$. 

\section{Single-particle inclusive}
\label{sec:Single-part-incl}

For definiteness we consider prompt photon
production, but the method sketched below
is more general \cite{lsv}.
A similar refactorization as in (\ref{eq:1})
holds for single-particle inclusive cross sections
at high $p_T$.
In particular, it contains the same $R_{f/f}$ functions.
The arguments supporting this refactorization are
somewhat more involved than for (\ref{eq:1}),
but reveal that only initial state radiation
contributes to $Q_T$ (the transverse momentum
of the $2\rightarrow 2$ parton cms frame)~\cite{LiLai}.
In the inclusive cross section, final-state interactions
require threshold resummation only. 
In contrast to electroweak annihilation, 
${\bf Q}_T$ is an unobserved variable, 
akin to $1-z$. Thus, the jointly resummed
prompt photon $p_T$ spectrum
may be written as an integral over ${\bf Q}_T$ of
a ``profile function'' \cite{lsv}
\begin{eqnarray}
  \label{eq:3}
  P_{ij}(N,{\bf Q}_T,Q) =
 \int\, d^2 \textbf{b} e^{-i \textbf{b}\cdot {\bf Q}_T}
e^{E_{ij \rightarrow \gamma k}}
\end{eqnarray}
where the exponential exhibits the joint resummation, as
\begin{eqnarray}
  \label{eq:4}
  {p_T^3 d \sigma^{({\rm resum})}_{AB\to \gamma} \over dp_T}
&=& 
\nonumber\\
&\ & \hspace{-19mm} \times
\sum_{ij} \frac{p_T^4}{8 \pi S^2} \int_{\cal C} {dN \over 2 \pi i}\;
\tilde{\phi}_{i/A}(N) \tilde{\phi}_{j/B}(N)
\nonumber\\
&\ & \hspace{-19mm} \times
\; \int_0^1 d\tilde x^2_T \left(\tilde x^2_T \right)^N\;
{|M_{ij}(\tilde x^2_T)|^2\over \sqrt{1-\tilde{x}_T^2}}\,
\nonumber\\
&\ & \hspace{-19mm} \times
C_\delta^{(ij\to \gamma k)}(\alpha_s,\tilde
x_T^2)
\int {d^2 {\bf Q}_T \over (2\pi)^2}\;
\Theta\left(\bar{\mu}-Q_T\right)
\nonumber \\
&& \hspace{-19mm} \times
\left( \frac{S}{4 {\bf p}_T'{}^2} \right)^{N+1} P_{ij}(N,{\bf Q}_T,Q)
\end{eqnarray}
with $\tilde{x}_T^2 = 4|\textbf{p}_T-{\bf Q}_T/2|^2/\hat{s}$,
$\tilde{\phi}_{i/A}(N)$ Mellin moments of the parton distributions,
$|M_{ij}|^2$ the Born amplitudes,
and $\bar{\mu}$ a cut-off restricting ${\bf Q}_T$ to sufficiently
small values for resummation to be relevant. 
The $C_\delta^{(ij\to \gamma k)}$ are infrared safe coefficient
functions, which include short-distance dynamics at the scale $Q$.
For notational simplicity 
we have suppressed all factorization and renormalization scale dependence.
Cross sections computed on the basis of (\ref{eq:4}) are shown in 
Fig.~\ref{fig:one}  as functions of $Q_T$ at fixed $p_T$.
The kinematics are those of the E706 experiment;~\cite{e706} see~\cite{lsv}
for details of the calculation, in particular those regarding the evaluation
of the $b$-integral in (\ref{eq:3}). The  dashed lines are 
$d\sigma^{(\rm resum)}_{{\rm pN}\to\gamma X}/dQ_Tdp_T$, with recoil 
neglected by fixing ${\bf p}_T'={\bf p}_T$, thus showing how each $Q_T$ 
contributes to threshold enhancement. The solid lines show the same, but 
now including the true recoil factor $(S/4{\bf p}_T'{}^2)^{N+1}$.
The resulting enhancement is clearly substantial. For small $p_T$, the
enhancement simply grows with $Q_T$, while for $p_T$ above 5 GeV it has a 
dip at about $Q_T = 5$ GeV, which remains substantially above zero. This 
makes it difficult to determine $\bar\mu$ on this basis alone.

So far, the numerical results given in~\cite{lsv} are primarily to be 
regarded as illustrations, rather than quantitative predictions. This applies
in particular to the resummed $Q_T$-integrated cross section, 
shown in Fig.~\ref{fig:two} for $p_T\ge 3.5$ GeV and $\bar\mu=5$ GeV. 
These figures demonstrate, however, the size of the additional
enhancement that recoil can produce, and its potential phenomenological 
impact. 

\begin{figure}
\epsfxsize190pt
\figurebox{130pt}{160pt}{fig1.epsi}
\caption{
$d\sigma_{{\rm pN}\to\gamma X}/dQ_Tdp_T$ at {\protect $\sqrt{s}=31.5$} GeV,
as a function of $Q_T$ for various values of photon $p_T$. Dashed lines
are computed without recoil ({\protect
${\bf p}_T'={\bf p}_T$} in (\ref{eq:4})), solid lines are with recoil.}
\label{fig:one}
\end{figure}

\begin{figure}
\epsfxsize190pt
\figurebox{130pt}{160pt}{fig2.epsi}
\caption{$Ed^3 \sigma_{{\rm pN}\to\gamma X}/dp^3$
for pN collisions at {\protect $\sqrt{s}=31.5$ GeV}. The dotted line
represents the full NLO calculation, while the dashed and solid lines
respectively incorporate pure threshold
resummation~{\protect \cite{thrphen}} and the joint resummation described
in this paper.  Data have been taken from~{\protect \cite{e706}}.}
\label{fig:two}
\end{figure}

\subsection*{Acknowledgements} 
The work of E.L.\ is part of the research program of the
Foundation for Fundamental Research of Matter (FOM) and
the National Organization for Scientific Research (NWO).
The work of G.S.\ was supported in part by the 
National Science Foundation,
grant PHY9722101. 
G.S.\ thanks Brookhaven National Laboratory for its hospitality.
W.V. is grateful to RIKEN, Brookhaven National Laboratory and the U.S.
Department of Energy (contract number DE-AC02-98CH10886) for
providing the facilities essential for the completion of this work.


\begin{thebibliography}{99}

\bibitem{cssrv}
J.C.\ Collins, D.E.\ Soper and G.\ Sterman, in {\it Perturbative
Quantum Chromodynamics},
ed.\ A.H.\ Mueller (World Scientific, Singapore, 1989).

\bibitem{Liunified}
H.-n.\ Li, {\it Phys.\ Lett.}\ {\bf B454}, {328}
(1999), hep-ph/9812363.

\bibitem{lsv} E.\ Laenen, G.\ Sterman and W.\ Vogelsang,
{\it Phys. Rev. Lett.}{\bf 84}, 4296 (2000), hep-ph/0002078.

\bibitem{lsvlong} E.\ Laenen, G.\ Sterman and W.\ Vogelsang,
hep-ph/0010080.


\bibitem{dyresum}
G.\ Sterman, {\it Nucl.\ Phys.}\ {\bf B281}, {310} (1987);
S.\ Catani and L.\ Trentadue, 
{\it Nucl.\ Phys.}\ {\bf B327}, {323} (1989),  {\bf B353},
183  (1991).

\bibitem{gath} J.G.M.\ Gatheral,
{\it Phys.\ Lett.}\ {\bf B133}, 9 (1983);
G.\ Sterman,  in: {\it Perturbative quantum
chromodynamics}, proc. of Tallahassee conference (1981), AIP Conference 
Proceedings, eds. D.\ W.\ Duke and J.\ F.\ Owens, p.22;
J.\ Frenkel and J.C.\ Taylor, {\it Nucl.\ Phys.}\ {\bf B246}, {231} (1984).

\bibitem{LiLai}
     H.-L.~Lai and H.-n.~Li, {\it Phys.\ Rev.}\ {\bf D58}, 114020 (1998),
hep-ph/9802414.

\bibitem{thrphen}
      S.~Catani, M.L.~Mangano, P.~Nason, C.~Oleari and
      W.~Vogelsang, JHEP {\bf 9903} (1999) 025, hep-ph/9903436.

\bibitem{e706} L.~Apanasevich {\it et al.}, E706 Collab., 
{\it Phys.\ Rev.\ Lett.} {\bf 81}, 2642 (1998), hep-ex/9711017.



\end{thebibliography}
\end{document}